\documentclass[superscriptaddress,twocolumn,showpacs,preprintnumbers,amsmath,amssymb]{revtex4}

\usepackage{graphicx}
\usepackage{dcolumn}
\usepackage{bm}

\begin{document}

\title{Dynamics of magnetization in frustrated spin-chain systems: Ca$_3$Co$_2$O$_6$}

\author{Yu.B.Kudasov}
\email{kudasov@ntc.vniief.ru}
\affiliation{Russian Federal Nuclear Center - VNIIEF, Mira str. 37, Sarov, 607188, Russia}
\affiliation{Sarov State Physics and Technology Institute, Dukhov str. 6, Sarov, 607188, Russia}
\author{A.S.Korshunov}
\affiliation{Russian Federal Nuclear Center - VNIIEF, Mira str. 37, Sarov, 607188, Russia}
\author{V.N.Pavlov}
\affiliation{Russian Federal Nuclear Center - VNIIEF, Mira str. 37, Sarov, 607188, Russia}
\author{D.A.Maslov}
\affiliation{Sarov State Physics and Technology Institute, Dukhov str. 6, Sarov, 607188, Russia}

\date{\today}
\begin{abstract}
The magnetization dynamics of the triangular lattice of Ising spin chains is investigated in the framework of 
a two-dimensional model.
The rigid chains are assumed to interact with the nearest neighboring chains, an external
magnetic field, and  a heat reservoir that causes the chains to change their states randomly with time. 
A probability of a single spin-flip process is assumed in a Glauber-like form. This technique allows describing
properly the steps in the magnetization curves
observed in Ca$_3$Co$_2$O$_6$ and their dependence on a magnetic field sweep rate and temperature. A transition 
from a low-temperature to high-temperature phase is also observed.
\end{abstract}


\maketitle

Ising spin-chains packed into a two-dimensional (2D) frustrated lattice demonstrate a complex magnetic behavior 
due to a combination of a low dimensionality and frustration. There are few groups of compounds in which the 
triangular lattice is formed by antiferromagnetic (AFM) Ising spin chains, e.g. CsCoCl$_3$, CsCoBr$_3$ \cite{mekata},
or ferromagnetic (FM) ones like Ca$_3$Co$_2$O$_6$ \cite{aas, drillon}. A spin-chain system Sr$_5$Rh$_4$O$_{12}$ discovered
recently has a complex magnetic structure of chains \cite{cao}. An interaction between the nearest neighboring chains
in the lattice is of the AFM type and much weaker than the intrachain one. 
However, it causes the frustration and a large variety of magnetic structures \cite{mekata, aas, drillon, cao, kudasovPRL, kudasovEPL}.

A step-like magnetization curve in Ca$_3$Co$_2$O$_6$ has drawn recently considerable attention \cite{hardy1,
drillon, hardy2, hardy3, maignan, kudasovPRL, kudasovEPL, petrenko, kageyama, fresard}. The number of the steps depends on a sweep rate of the
external magnetic field and temperature \cite{drillon, hardy2, maignan}. Two steps become apparent in the temperature
range from 12~K to 24~K \cite{maignan}. The first step takes place at the zero magnetic field. Then a plateau at 
about $1/3$ of the full magnetization stretches up to the magnetic field of 3.6~T where the second
step to the saturated FM state occurs.  At least four equidistant steps are clearly visible below 12~K at a
moderate magnetic field sweep rate \cite{drillon, hardy2}. At an extremely low sweep rate the magnetization curve becomes close to the two-step
shape, similar to that is observed at the high temperatures \cite{hardy2}. A response to alternating magnetic fields and its dependence on the temperature 
and magnetic field were investigated carefully in Ref.\cite{hardy1}. The experimental results show that there exist 
two characteristic time scales of the magnetization dynamics. The first is of order of 1~s and the second reaches 
a few hours.

A crystal structure of Ca$_3$Co$_2$O$_6$ consists of Co$_2$O$_6$ chains running along the $c$ axis. The Ca ions are situated
between them and are not involved in magnetic interactions. The chains are made up of alternating face-sharing
CoO$_6$ trigonal prisms and CoO$_6$ octahedra. The
crystalline electric field splits the energy level of Co$^{3+}$ ions into the high-spin ($S=2$) and low-spin ($S=0$)
states. The chains form triangular lattice in the
$ab$ plane that is perpendicular to the chains. An in-chain exchange interaction between high-spin cobalt ions 
through the octahedra with low-spin cobalt ions is ferromagnetic. The parameter of the FM in-chain coupling
($J_F$) was found from the magnetic susceptibility at high temperatures \cite{kageyama}, specific heat 
\cite{hardy3}, and theoretical calculations \cite{fresard}. These estimations are in a reasonable agreement with 
each other ($J_F \approx 25$K). The weak AFM interchain interaction causes the partially disordered AFM (PDAFM) structure 
of chains (or honeycomb magnetic structure) below $T_{C1}$=24~K. At $T_{C2} \approx 12$~K another magnetic transition takes place \cite{hardy3, petrenko}. It should be mentioned that the topology of the magnetic net in Ca$_3$Co$_2$O$_6$ is rather complex, e.g. 
there exist helical paths \cite{fresard}.

\begin{figure*}
\includegraphics{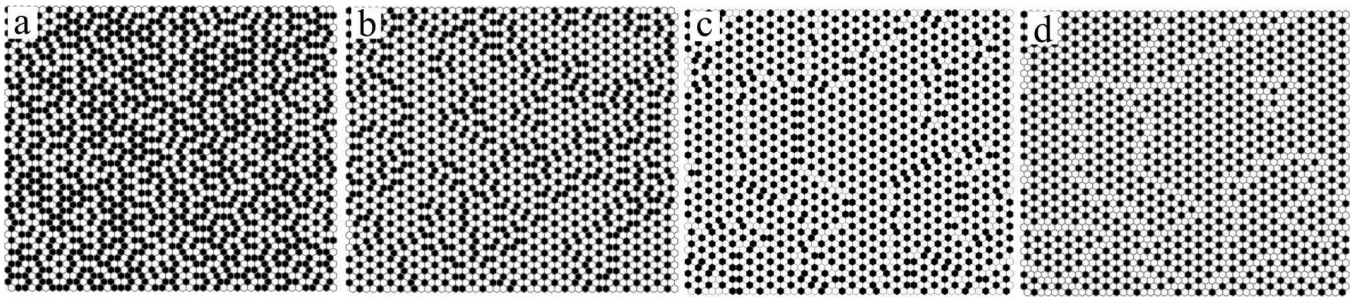}
\caption{\label{f1} Fragments of the triangular lattice during the simulation at $T=4$~K and 1~T/min (from left to right): 
the initial state $B=0$~T, $B=0.768$~T, $B=1.824$~T, and $B=2.784$~T. 
The white and black circles stand for spin-up and spin-down chains.}
\end{figure*}

Taking into account the strong intrachain interaction one can assume the chains to be in two ordered states 
(spin-up or spin-down) at low temperatures. This is a basic assumption of a rigid-chain model \cite{kudasovPRL}. Then the problem 
is reduced to the AFM triangular Ising model \cite{wannier, kudasovPRL}:
\begin{equation}
H=J\sum_{<ij>}{\sigma_i \sigma_j}-B\sum_{i}{\sigma_i}
\label{Ising}
\end{equation}
where $\sigma_i=\pm{1}$ is the $c$-axis projection of the $i$-th chain spin, $J>0$ is the parameter of the AFM
interchain coupling, $B$ is the external magnetic field, $<ij>$ denotes the summation over all the nearest-neighbor pairs on 
the triangular lattice. 

A second assumption of the
rigid-chain model is that even at a very low magnetic field sweep rate the system is out of equilibrium, i.e. it is in a metastable state rather 
than in the ground state \cite{kudasovPRL}. 
The large hysteresis loop and strong
dependence of magnetization curve on the magnetic field sweep rate favor this point. Conditions of the metastability of the system 
can be formulated in the following form \cite{kudasovPRL}
\begin{equation}
\sigma_i h_i\leq 0
\label{meta}
\end{equation}
where $h_i = J\sum_{<ij>}{\sigma_j}-B$ is the effective field for the $i$-th chain. This inequality should be satisfied for all the chains.
Following the single-flip
technique for the nonequilibrium Ising model \cite{kim} the probability $W_i$ of a spin-flip event at the $i$-th chain was formulated as
\begin{eqnarray}
W_i=\left \lbrace \begin{array}{cc} 
0 & \text{ if }\sigma_i h_i < 0, \\ 
1 & \text{ if }\sigma_i h_i \geq 0. 
\end{array}\right.
\label{W1}
\end{eqnarray}
The rigid-chain model describes properly the four-step magnetization curve in  Ca$_3$Co$_2$O$_6$. 
On the other hand one should keep in mind that once the system occurs in a metastable state it will dwell in 
this state for an arbitrarily long time, i.e. this is a frozen metastability approximation.

Recently, static magnetization curves in Ca$_3$Co$_2$O$_6$ were investigated by means of Monte Carlo technique 
\cite{yao}. It was observed that the perfect triangular lattice of the rigid spins produces the two-step magnetization 
curve even at low temperatures. To overcome this problem the nearest-neighbor
interactions were radomized. The magnetization curves demonstrated the four-step behavior on such imperfect lattice.
However, it should be pointed out that these two types of the magnetization curves were observed experimentally on 
the \emph{same} sample at different sweep rates \cite{hardy2}. Therefore the four-step magnetization curve can 
not originate from the lattice imperfection.     

In the present paper, we generalize the rigid-chain model to include dynamical aspects of the problem. Following to
the Glauber theory \cite{glauber} we assume that the chains interact not only with the nearest neighbors and external
magnetic field but also with a heat reservoir. In this case the probability of a spin flip of the $i$-th chain per time unit can be written down as
\begin{equation}
W_i=\frac{\alpha}{2} 
 \left[1-\sigma \cdot \tanh \left( - \frac{J}{kT}\sum_{<ij>} \sigma{_j} + \frac{\mu B}{kT}\right) \right]  
\label{Glauber}
\end{equation}
where $\alpha$ is the constant of the interaction of a chain with the heat reservoir, $k$ is the Boltzmann constant, 
$T$ is the temperature, $\mu$ is the chain magnetization. It should be mentioned that the ratio $J/\mu$ do not 
depend on the chain length. It is determined by the plateau length in the magnetization curve 
($\Delta B= 1.2$~T for Ca$_3$Co$_2$O$_6$). In effect a temperature behavior of the system is related to the 
effective temperature $kT/N$ where $N$ is the length of the chain (number of magnetic ions).

One can see that if $T \rightarrow 0$ the expression (\ref{Glauber}) goes to
the function (\ref{W1}), i.e. the model of Ref.\cite{kudasovPRL} is the low-temperature limit of the present one. 
In the opposite case $T\rightarrow\infty$ the probability of a spin-flip event is
independent on the magnetic field and interchain coupling. At finite temperatures the 
probability depends on the ratio of the effective field to the temperature and is nonzero even if $\sigma_i h_i < 0$.
That is why, an evolution of metastable states exists within this model. 

We have performed a numerical simulation of the magnetization evolution on a rhombic $96{\times}\, 96$ supercell of the triangular 
lattice with periodic boundary conditions. Few periodic structures can appear during the simulation.
To avoid artificial domain boundaries the length of the supercell should be multiple of a periodicity of these structures. In order to satisfy this condition for any three- or two-sublattice ordering the length was chosen to be multiple of 6. Preliminary calculations on $24{\times}\, 24$, $48{\times}\, 48$ , and $192{\times}\, 192$ supercells showed a good convergence of the results. The best coincidence of the theoretical and experimental data occurred at $\alpha = 0.21$~s$^{-1}$
and $N=30$ which are reasonable values. They were used in all further calculations. 

The simulation started with preparation of an initial state. As it was discussed in Ref.\cite{kudasovPRL} the PDAFM
structure is rather close to the ground state of the AFM triangular Ising model. That is why, the initial state 
was obtained by an evolution of the PDAFM phase at the zero magnetic field. When the duration of this preliminary evolution was more than
$5\alpha^{-1}$, results of the simulation became practically independent on the initial state. The typical duration of the preliminary evolution in
our calculations was about $100\alpha^{-1}$. A fragment of the initial state is shown in Fig.\ref{f1}a. It should be mentioned that one can 
prepare the initial state from an arbitrary random state. However it requires a longer period to get the initial state.

\begin{figure*}
\includegraphics[width=0.49\textwidth]{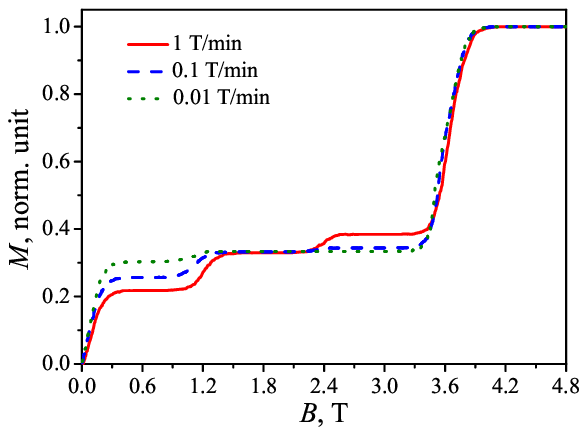}
\includegraphics[width=0.49\textwidth]{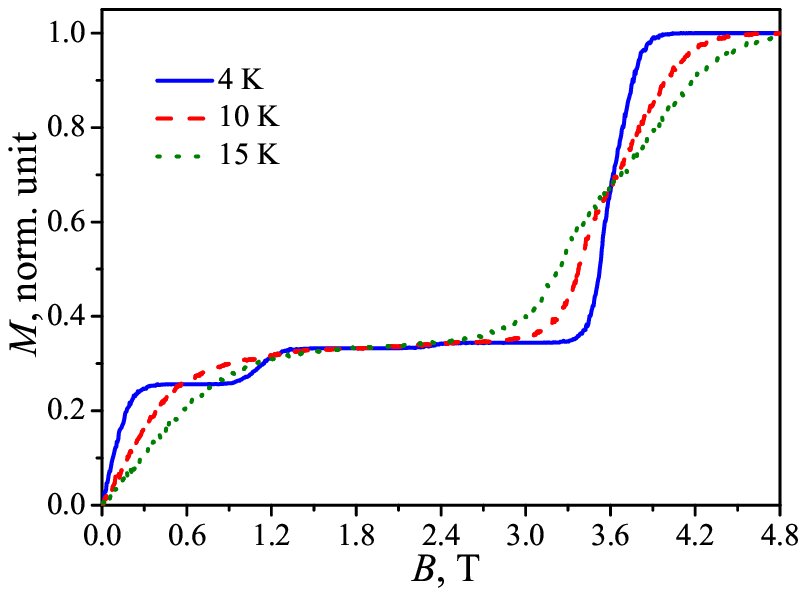}
\caption{\label{f2} The magnetization curves in the increasing magnetic field at different sweep rates and $T=4$~K (left) and at different 
temperatures with the constant magnetic field field sweep rate of $0.1$~T/min (right).}
\end{figure*}
%

After obtaining the initial state the magnetic field was switched on at a constant
sweep rate. We adjusted the magnetic field sweep rates to the experimental values of Ref.\cite{hardy2}. A typical number of time steps was about $2\cdot10^6$ per a full cycle (increasing-field and decreasing-field branches).

Fig. \ref{f2} demonstrates an average magnetic moment as a function of magnetic field $M(B)$ calculated for different magnetic field sweep rates at a constant temperature (in the left panel) and for different temperatures at a constant sweep rate (in the right panel). One can see that the second and third steps in the magnetization curve vanish with decreasing of the magnetic field sweep rate or with increasing the temperature. This general tendency and shape of the magnetization curves are in a very good agreement with the experimental observations \cite{hardy2}. 

To clarify an origin of such behavior
we present specimens of configurations at different magnetic fields corresponding to the plateaus in the magnetization curves at the highest 
magnetic field sweep rate (see Fig.\ref{f1}). One can clearly see domains in the figures. The ferrimagnetic state is
formed by three sublattice. It is obviously three-fold degenerate. While the magnetic field sweep rate is sufficiently fast
the ferrimagnetic phase starts growing at a number of nucleation centers. This causes the domain formation. 
With a decrease of the sweep rate the domains expand and for the extremely slow sweep rate the domain walls almost disappear. 
It is easy to see that the single-domain ferrimagnetic state leads to the two-step magnetization curve because all the spin-down chains are encircled by six spin-up chains. The dotted curve in the left panel of Fig.\ref{f2} corresponding to the extremely low magnetic field sweep rate is similar to that was obtained by the Monte Carlo technique \cite{yao}. 

\begin{figure*}
\includegraphics[width=0.49\textwidth]{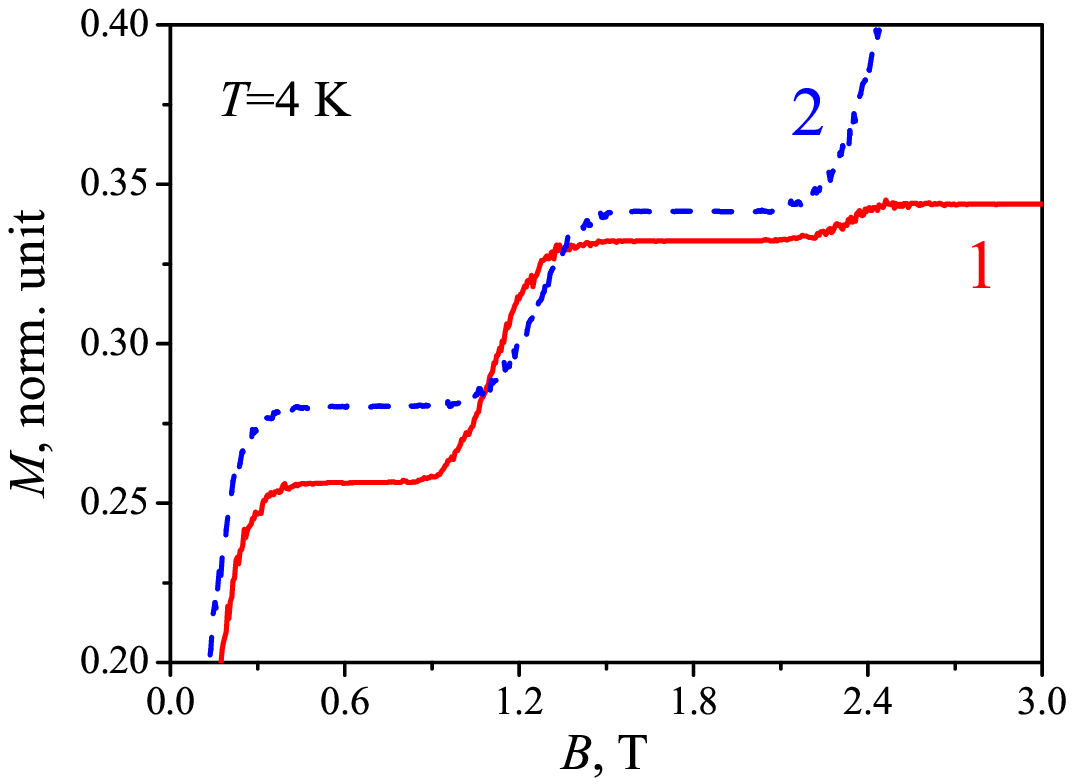}
\includegraphics[width=0.49\textwidth]{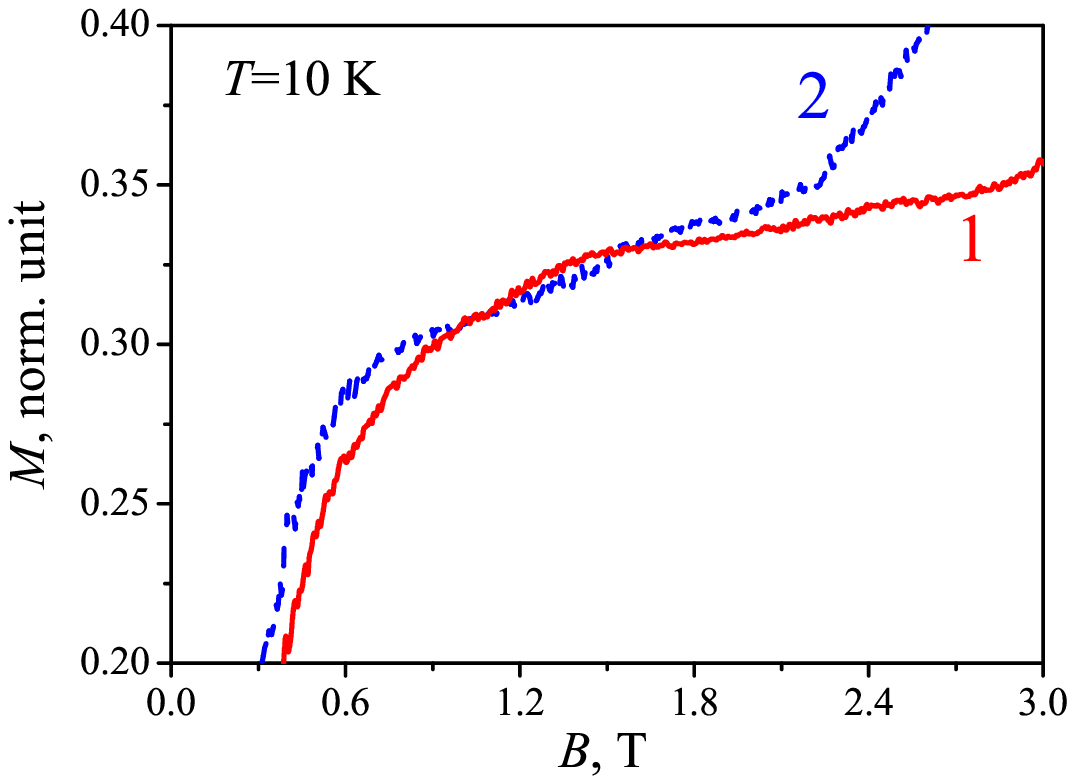}
\caption{\label{f3} Enlargements of hysteresis loops with a sweep rate of $0.1$~T/min at $4$~K and $10$~K. 
The solid lines and dashed lines are the field-increasing and field-decreasing branches, respectively.}
\end{figure*}

%
Domain boundaries contain chains with other environments and cause appearance of the two additional steps in the magnetization curve. 
For instance, at the first plateau (Fig.\ref{f1}b) there are spin-down chains with 2 spin-down and 4 spin-up nearest neighbors. 
Their critical spin-flip field is 1.2~T. At the second plateau (Fig.\ref{f1}c) there exist spin-down chains with 1 spin-down and 5 spin-up nearest 
neighbors. Their critical spin-flip field is 2.4~T. The higher the magnetic field sweep rate is, the smaller the domains become and the more 
apparent the additional magnetization steps are. 

Here we should mention that the chains in the domain boundaries as well as the internal chains are in metastable states at the plateaus, that is, they are 
oriented along the effective fields. The spin-flip probability in the form of Eq.(\ref{Glauber}) allows some fraction of exited states that, 
in their turn, leads to a creep of the domain wall. This is a new important ingredient which appears in the present theory. It explains
the existence of two characteristic times in Ca$_3$Co$_2$O$_6$. The first one is related to a single spin-flip evens and specifies the
fast processes of about 1~s. The second is due to the creep of the domain boundaries (up to 10$^4$~s).

As it was argued in Ref.\cite{kudasovEPL} the transition from the low-temperature to high-temperature regime stems from disordering of a fraction
of the chains. This mechanism is essentially three-dimensional. That is why, the observation of this transition within the
present 2D model is surprising. The present interpretation of the transition differs from that was assumed in Ref.\cite{kudasovEPL}. 
It arises due to a drastic increase of a domain boundary mobility at high temperatures that leads to establishing of the 
single-domain ferrimagnetic state during a short time.

Another interesting features that were observed experimentally are crossings of field-increasing and field-decreasing branches of the
magnetization curve \cite{hardy2}. As one can see in Fig.\ref{f3} they are also reproduced by the simulations.

In conclusion, in the present paper we go beyond the frozen metastability approximation of Ref.\cite{kudasovPRL}. The simulation of the nonequilibrium evolution was performed by means of a Glauber-like form of the spin-flip probability 
that allows us investigating the dependence of the magnetization curves on the temperature and magnetic field sweep rate in a good
agreement with the experimental data. At the extremely low magnetic field sweep rate the results of the 2D simulation are similar to that
was obtained by the Monte Carlo
technique \cite{yao}.
The present model gives reasonable values of both the critical temperatures: $T_{C2} \approx 10$~K and $T_{C1} \approx 20$~K. Although it is not
valid at high temperatures ($T \gtrsim 15$~K) since the intrachain disorder becomes essential \cite{kudasovEPL}.
An experimental feature that escapes the 2D simulation is a drastic increase of the hysteresis 
loop at a very low temperature ($T=2$~K) as compared to moderate temperatures ($T \sim 5$~K). Most probably it is related to an intrachain
dynamics of the magnetization and requires a 3D simulation to be describe properly. 

The work was supported by ISTC (project \#~3501), RFBR (projects 08-02-97018-r-povolzhje-a, 08-02-00508-a), and Federal Agency of Science
and Innovations of Russian Federation. D.A.M. acknowledges a support of the ``Dynasty'' foundation.


\end{document}